\documentclass[acmsmall, screen, nonacm]{acmart}

\usepackage[nodayofweek]{datetime}
\usepackage{amsmath, amsfonts, bbm}
\usepackage{dsfont}
\usepackage{longtable}
\usepackage{multirow}
\usepackage{listings}
\usepackage{hyperref}
\usepackage{subcaption}
\usepackage{lineno}
\usepackage{nomencl}
\usepackage[lined,boxed]{algorithm2e}

\begin{document}

% \begin{frontmatter}

% \title{A quantum annealing application for fairness in heat grid conceptual designs}
\title[DHN fairness]{Fairness evaluation during the conceptual design of heat grids with quantum annealers}
\author{Kelvin Loh}
\email{kelvin.loh@tno.nl}
\orcid{0000-0002-8085-6495}
\affiliation{
    \institution{TNO - Quantum Technology}
    \streetaddress{Stieltjesweg 1}
    \city{Delft}
    \postcode{2628CK}
    \country{the Netherlands}
}

\begin{abstract}
This paper presents a workflow to evaluate the macro scale thermal fairness for producers in a district heating network during the conceptual design phase of such a network. It uses the workflow to evaluate two types of proposed topologies for a future network to be constructed in Europe. The workflow is also novel in the sense that it has been demonstrated in the paper that within the implementation, the load balancing step can be solved readily by a quantum computing system (in particular, a quantum annealer from DWave). The concept of fairness is also addressed in the workflow and the results show that there exist an optimum number of producers for a given district heating topology.
\end{abstract}



% \begin{keyword}

% \end{keyword}

% \end{frontmatter}

% \begin{titlepage}
% \end{titlepage}

% \signaturepage{}

\maketitle

\newpage
\makenomenclature
\nomenclature{QUBO}{Quadratic Unconstrained Binary Optimization}
\nomenclature{QPU}{Quantum Processing Unit}
\nomenclature{QA}{Quantum Annealing}
\nomenclature{KPI}{Key Performance Index}
\nomenclature{DH(N)}{District Heating (Network)}
\nomenclature{4GDH}{4th Generation District Heating}
\nomenclature{NP problem}{Nondeterministic Polynomial problem}
\nomenclature{NISQ}{Noisy Intermediate Scale Quantum device}
\nomenclature{VQE}{Variational Quantum Eigensolver}
\nomenclature{QAOA}{Quantum Approximate Optimization Algorithm}
\printnomenclature

\newpage
\section{Introduction}
\label{Chap:Introduction}
The concept of using district heating (DH) networks as a method of providing heat and hot water to households are gaining popularity within Europe. In the Netherlands, there are now increased initiatives at the local municipalities~\cite{ecn:DHNNL:2015} and federal government levels to replace gas networks with 4th generation district heat grids (4GDH). The idea is to use multiple sources of hot water (such as from geothermal wells, and waste heat from industrial plants) to provide for the consumers within the network, thus, reducing or replacing the need of gas as the main source of heating energy. There have already been many investigations done on the characteristics of a 4GDH (\cite{REZAIE20122}, \cite{RePEc:eee:energy:v:68:y:2014:i:c:p:1-11}, \cite{LUND20101381}). Amongst the key characteristic is that 4GDH involves strategic and innovative planning and the integration of DH into the operation of smart energy systems~\cite{LUND2018147}. This means that when planning and designing for a DH, (e.g. during the conceptual design phase) network designers need to take a holistic approach to come up with robust designs for all the stakeholders within the DH network. There have been many works on optimization of DH during the design~\cite{LiDHNdesign:2013} and operation~\cite{SAKAWA2001139} phases, however, none has taken into account the concept of fairness during the design phase~\cite{ioan:2019}.

This paper investigates on the idea of thermal fairness~\cite{Bhattacharya:2016:TDR:2934328.2934343} (in the reference, micro-level fairness is applied to the consumers instead of producers of the network during operation planning to provide a fair demand response), which on a macro scale could correspond to a measure of the tendency for monopolistic business positions to occur within the network. We claim that the topology of the network itself combined with the consumer demands in the network would affect the macro fairness of the network (taking the concept of fairness from computer networks~\cite{fairness_computer_networks:2000}), thus either promoting or discouraging monopoly. The workflow and implementation presented here could be used in the conceptual design phase for the network designer to encourage fair and even competition within a DH network. We apply the concept of fairness to an upcoming DHN in Europe and use actual energy demands of the consumers in the proposed network in our approach. This paper also demonstrates the use of a quantum annealer to solve the generic $k$-graph partitioning problem when we need to split the demands of the consumers fairly amongst the total number of producers within the grid. We thus claim that the solution to a balanced $k$-graph is fair for the producers and policy makers can use this as a measure to either reward or penalize business practices which deviate from the optimum solution.

\subsection{DHN Fairness}
Typically in the design for 4GDH networks, a monopolistic position of the DHN operator is always assumed~(\cite{BUFFA2019504}, \cite{cma:heatnetworks:2018}). This is a problem often mentioned in the literature~\cite{osman:monopoly_business:2015}. The reason this occurs is because the owner of the network is also responsible for heat delivery, thus providing no room for competition. In the Dutch heat market where customers are used to switching energy providers (through electricity and natural gas networks), the monopolistic position of DHNs is perceived as a step backwards in terms of fair competitions. In order to protect the customers of DHNs from the monopolistic position of DHN operators, the government established the Dutch Heat Act~\cite{osman:monopoly_business:2015}. In order to improve upon this situation further, we need to already include the concept of fairness during the conceptual design phase of a 4GDH network~\cite{cma:heatnetworks:2018}. This approach is as recommended by the Competition and Markets Authority in the UK, recommending that network design and build should be improved to better align the interests of the stakeholders in the network~\cite{cma:heatnetworks:2018}.

\subsection{Quantum Technology}
\label{subsec:QC}
As we are using a relatively new technology to solve the classical combinatorial optimization problem, we briefly mention about the technology. There are two types of quantum computing implementations~\cite{2018arXiv180100862P}. The first is what we know as a universal gated quantum computer and the other is a quantum annealer. 

We can solve our problem in gated quantum computers using an algorithm known as Quantum Approximate Optimization Algorithm (QAOA). This is a hybrid optimization algorithm introduced by~\cite{2014arXiv1411.4028F} to take advantage of Noisy Intermediate Scale Quantum (NISQ)~\cite{2018arXiv180100862P} devices to demonstrate that early quantum computers can already be useful for certain classical combinatorial problems. However, the number of qubits required to solve our problem is still too large compared to current publicly available gated quantum computers. As such, in this paper, we focus on using the (DW\_2000Q) quantum annealer system provided by DWave~\cite{DWave:git:2017}.

Quantum annealing (QA) is the quantum counterpart to simulated annealing. It takes advantage of quantum mechanical fluctuations instead of thermal, to get to an optimal solution with faster convergence~\cite{kadowaki:1998}. The D-Wave system minimizes the following Quadratic Unconstrained Binary Optimization (QUBO) objective function (equation~\eqref{eq:qubo}).

\begin{align}
    \label{eq:qubo}
    O(\mathbf{Q}, \mathbf{x}) &= \sum_i Q_{ii} x_i + \sum_{i < j} Q_{ij}x_i x_j
\end{align}

By tuning the input of the objective function $\mathbf{Q}$, we change the system Hamiltonian for the qubits. The output of the anneal is a low-energy ground state, which consists of an Ising spin for each qubit in the system. This is the mechanism upon which a quantum annealer is constructed which gives the ability to quickly solve certain classes of NP-hard complex problems such as optimization, machine learning and sampling problems~\cite{2017arXiv170503082U}.

As we are trying to solve a graph partition problem in the energy domain, we are only aware of the work~\cite{AJAGEKAR201976} in energy systems optimization using quantum computing systems.

\section{Theoretical Background}
\label{Chap:Theory}
In this section we describe the theoretical foundations of our implementation. When we want to evenly distribute a set of nodes in a given network, the problem is known as a graph partitioning problem. We start by formulating this problem to a QUBO problem for $k$-clusters~\cite{2017arXiv170503082U} such that we can solve this problem with the DWave system.

Let $G(V,E)$ be the graph which connects the consumers where the nodes are the consumer nodes and edges represent the proposed connection between the consumer nodes. We would like to partition this graph between $k$ producers. The network has $n$ nodes and $m$ edges. Note that the producers are assumed to have infinite capacities and are always capable of meeting the demands of the network. Another assumption is that the producers are not fixed in any location.

We start the theoretical foundation by first deriving the QUBO formulation of our problem.

\subsection{Derivation of the QUBO}
\label{subsec:DerivationQUBO}
In order to partition a graph concurrently into an arbitrary number of parts, $k$, we need to formulate the problem as a QUBO, which takes on binary variables. The decision variables are given by
\begin{align}
    x_{i,j} &=  \begin{cases}
                    1,& \text{if node } i \text{ is in producer } j\\
                    0,              & \text{otherwise}
                \end{cases}
\end{align}
The constraint
\begin{align}
\label{eq:constraint_one}
    \sum_{j=1}^k x_{i,j} &= 1
\end{align}
for each node $i$ ensures that each node is in exactly one producer.
The second constraint
\begin{align}
\label{eq:constraint_fair}
    \sum_{i=1}^n x_{i,j} &= \frac{n}{k}
\end{align}
for $j = 1,...,k$ are the balancing constraints for the producer sizes.
\begin{align*}
    \mathbf{x}_j = \left[ 
                    \begin{matrix}
                    x_{1,j}\\
                    x_{2,j}\\
                    \vdots\\
                    x_{n,j}
                    \end{matrix}
                    \right]
\end{align*}
The number of cut edges across the $k$ producers is given by
\begin{align}
\label{eq:min_cut_equal}
    \frac{1}{2} \left( \sum_{j=1}^k \mathbf{x}_j^\text{T} L \mathbf{x}_j \right)
\end{align}
where $L$ is the Laplacian matrix of the graph.

The cost function is then given by
\begin{align}
    \label{eq:cost_function}
    \beta \left( \sum_{j=1}^k \mathbf{x}_j^\text{T} L \mathbf{x}_j \right) + \sum_{j=1}^k \alpha_j \left( \sum_{i=1}^n x_{i,j} - \frac{n}{k} \right)^2 + \sum_{i=1}^n \gamma_i \left( \sum_{j=1}^k x_{i,j} - 1 \right)^2
\end{align}
where $\beta$, $\alpha_i$, and $\gamma_i$ are positive penalty constants.

Following from the derivation and the variables as presented by~\cite{2017arXiv170503082U}, the QUBO formulation is
\begin{align}
    min_{\mathbf{x}} \left( \mathbf{X}^\text{T} (\beta \mathcal{L} + \alpha \mathcal{I} + \mathbf{B}_\Gamma) \mathbf{X} - ( 2\Gamma^\text{T} + 2\frac{n}{k}\alpha \mathbbm{1}_{\mathcal{N}}^\text{T} )\mathbf{X}   \right)
\end{align}
where $\mathbf{X}^\text{T} = \left[ \mathbf{x}_1^\text{T} , \dots, \mathbf{x}_k^\text{T} \right]$.

As we have noticed, Equation~\eqref{eq:cost_function} assumes that the demands are equally weighted and the edge weights are also equal. We can modify the problem in such a way that we fulfil different demand constraints and different distances between the consumer nodes.
In order to take into account the different demand constraints, we need to modify the constraint on the equal parts Eq~\eqref{eq:constraint_fair} to
\begin{align}
\label{eq:fair_weighted}
    \sum_{i=1}^n w_i x_{i,j} &= \frac{\sum_{i=1}^n w_i}{k}
\end{align}
where $w_i$ is the weighted demand of the $i$-th consumer.
As for the condition that each cluster should minimize the total distances between the member nodes, we can modify Equation~\eqref{eq:min_cut_equal} to
\begin{align}
\label{eq:min_cut_min_distance}
    \frac{1}{2} \left( \sum_{j=1}^k \mathbf{x}_j^\text{T} \Delta_{\mathcal{L}} \mathbf{x}_j \right)
\end{align}
where $\Delta_{\mathcal{L}}=$ off-diagonal elements of the matrix $(-I D I^\text{T})$, $I \in \mathbb{R}^{n \times m}$ is the incidence matrix of the network, $D \in \mathbb{R}^{m \times m}$ is a diagonal matrix with entries $d_{i,i} = {i\text{-th edge distance}}$.

The modified cost function is then given by Equation~\eqref{eq:modified_cost_function}.

\begin{align}
    \label{eq:modified_cost_function}
    \beta \left( \sum_{j=1}^k \mathbf{x}_j^\text{T} \Delta_{\mathcal{L}} \mathbf{x}_j \right) + \sum_{j=1}^k \alpha_j \left( \sum_{i=1}^n w_i x_{i,j} - \frac{\sum_{i=1}^n w_i}{k} \right)^2 + \sum_{i=1}^n \gamma_i \left( \sum_{j=1}^k x_{i,j} - 1 \right)^2
\end{align}

\subsection{Fairness measures and Heat Grid overall KPI}
\label{subsec:FairnessMeasures}
In order to evaluate whether a network design is capable of being fair and also to judge whether the optimized solution of equation~\eqref{eq:modified_cost_function} is indeed fair, we need to first define what we mean by fairness of a network, and how it applies to the heat grid with a customized KPI. We consider first the equal distribution of demands in our KPI. For a system which allocates the demands to $k$ producers such that the $i$-th producer needs to provide $y_i$ of the total network demand, then the intuitive measures for fairness given a demand distribution would be through the variance, coefficient of variation or the min-max ratio \cite{Jain:1984:QMF}. In this paper, we would use the Jain fairness index, equation~\eqref{eq:jain_fairness_index} as proposed by \cite{Jain:1984:QMF}.

\begin{align}
    \label{eq:jain_fairness_index}
    f_{\text{jain}} (y) &= \frac{\left [ \sum_{i=1}^k y_i \right]^2}{k \sum_{i=1}^k y_i^2}
\end{align}

The index, $f_{\text{jain}}(y) \in \mathbb{R}_{[0,1]}$, has the value of 1 when the total network demands are being equally distributed amongst the producers (thus, the solution is 100\% fair). As the disparity increases, the fairness index decreases, and the solution which favors a few producers would be indicated by an index which is close to 0.

In addition to the demand being equally distributed, we need to take into account the distances of the pipes connecting the demand nodes to each producer. Ideally, each producer groups should have minimal pipe lengths (i.e. minimal edge lengths) connecting the demand nodes so as to minimize transportation costs due to a reduction in heat loss. Thus, we also need to evaluate the distances term which contributes to the KPI. We let $d_{i,j}$ as the physical distance of the $i$-th path, assigned to the $j$-th producer. In other words, the $i$-th path is the shortest path which connects two nodes that are assigned to producer $j$. We then come up with equation~\eqref{eq:distance_index} to evaluate the quality of the solution based on the distance criteria alone.
\begin{align}
    \label{eq:distance_index}
    D &= 1 - \frac{\sum_{j=1}^{k} \sum_{i=1}^{n_k} d_{i,j}}{\sum_{i=1}^{N_c} d_{i}}
\end{align}
where $n_k$ is the total number of node combinations assigned within the producer set $k$, and $N_c$ is the total number of path combinations for a complete graph consisting of the demand nodes. The distance index is scaled by the maximum possible distance covered by all possible combinations of paths so that the resulting index, $D \in \mathbb{R}_{[0,1]}$. In effect, smaller clusters within each producer groups would cause the value of $D$ to tend to 1.

Since both $y$ and $d$ from equations~\eqref{eq:jain_fairness_index} and~\eqref{eq:distance_index} depend on the solution $X$ of equation~\eqref{eq:modified_cost_function}, the KPI to evaluate the solution for a given network is then simply given by equation~\eqref{eq:KPI_equation}
\begin{align}
    \label{eq:KPI_equation}
    KPI \left(\mathbf{X} \right) &= \alpha_j \cdot f_{\text{jain}} \left(\mathbf{X} \right) + (1 - \alpha_j) \cdot D \left(\mathbf{X} \right)
\end{align}
where $\alpha_j \in \mathbb{R}_{[0,1]}$ represent the relative importance factor between the KPI terms. Throughout the paper, the presented cases use an equal weighting (i.e. $\alpha_j = 0.5$) unless stated otherwise.

\section{Implementation}
\label{Chap:Implementation}
In this section we describe the implementation in more detail. First we start by describing how we determine the weighting factor of each node as they need to represent the actual heat demands inside the proposed network. After that we present the two proposed networks for which we need to evaluate their fairness indices with different number of producers in the network. We next discuss the solvers for which we can obtain the optimized solutions of equation~\eqref{eq:modified_cost_function}.

\subsection{Demand nodes}
\label{subsec:weighting_demands}
In order for us to represent the actual heat demands in the network, we start by transforming the energy demands profile of the year 2015 (with $Nt$ discrete number of timesteps) for each node to a static form. We take the max of the annual demand of each node. This is because the network has to be able to handle the worst case scenario of providing for the maximum energy demands from the users of the network. We take the max of the annual demand for each node as it represents the highest possible energy demand which a particular node would expect to be delivered by the network producers. We then scale it by the sum over all the consumers since we would like to have a weight factor which sums to 1. Let $w_{i,t}$ be the time-dependent demand profile of the $i$-th consumer node, at timestep $t$ of the profile. The fraction used as input when solving for equation~\eqref{eq:modified_cost_function} is then given by equation~\eqref{eq:demand_weightage}.

\begin{align}
    \label{eq:demand_weightage}
    w_i &= \frac{\max_{t \in Nt} \left( w_{i, t} \right) }{\sum_{i=1}^{n} \max_{t \in Nt} \left( w_{i, t} \right) } 
\end{align}

\begin{figure}
    \centering
    \begin{subfigure}[b]{0.45\textwidth}
        \includegraphics[width=0.95\linewidth]{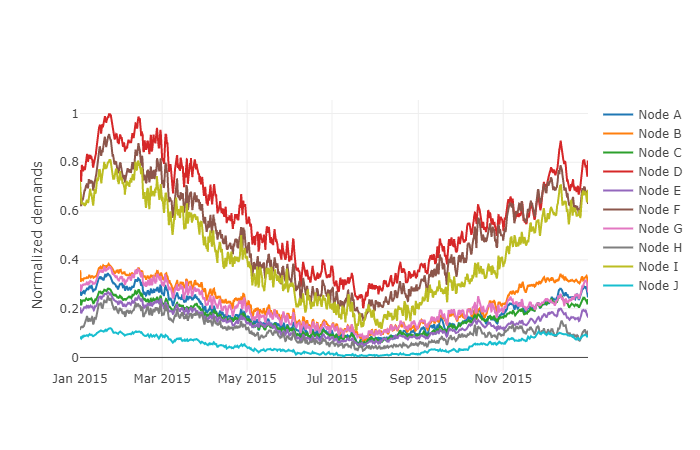}
        \caption{Normalized energy demands per node for 2015}
        \label{fig:norm_energy_demands}
    \end{subfigure}
    ~ %add desired spacing between images, e. g. ~, \quad, \qquad, \hfill etc. 
    \begin{subfigure}[b]{0.45\textwidth}
        \includegraphics[width=0.95\linewidth]{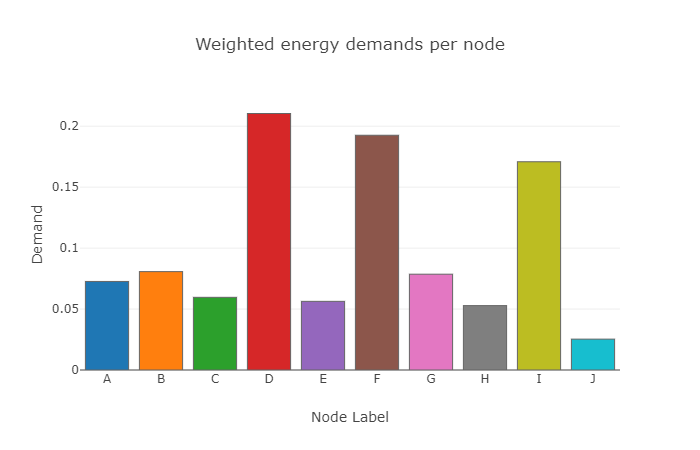}
        \caption{Weighted energy demands per node for 2015}
        \label{fig:weighted_energy_demands}
    \end{subfigure}
    \caption{Demands profile per node}
    \label{fig:demands_per_node}
\end{figure}

The effect of applying equation~\eqref{eq:demand_weightage} to the demand profiles is shown in figure~\ref{fig:demands_per_node}. The normalized profile in figure~\ref{fig:norm_energy_demands} is translated to figure~\ref{fig:weighted_energy_demands} for each consumer nodes.

\subsection{Topology types}
\begin{figure}[htbp]
    \centering
    \begin{subfigure}[b]{0.45\textwidth}
        \includegraphics[width=0.95\linewidth]{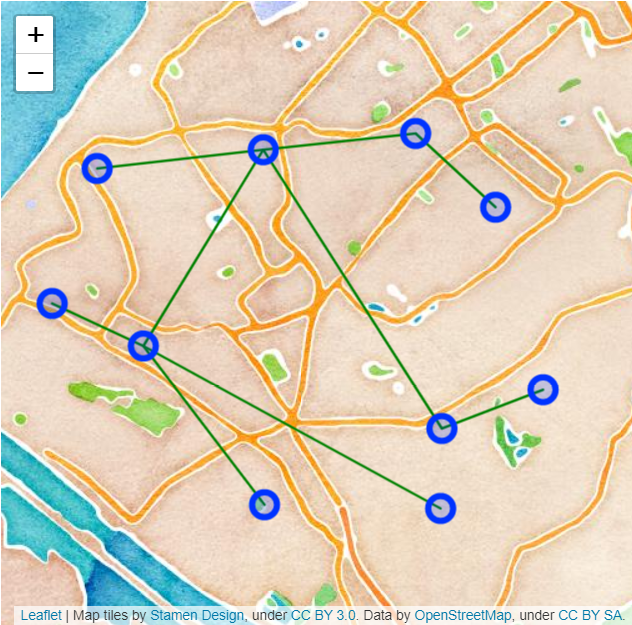}
        \caption{Tree-like topology}
        \label{fig:tree_topo}
    \end{subfigure}
    ~ %add desired spacing between images, e. g. ~, \quad, \qquad, \hfill etc. 
    \begin{subfigure}[b]{0.45\textwidth}
        \includegraphics[width=0.95\linewidth]{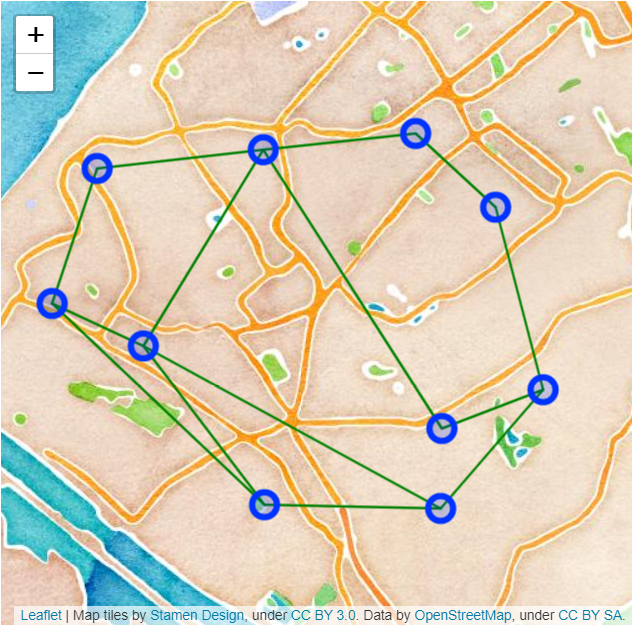}
        \caption{Circular-like topology}
        \label{fig:circular_topo}
    \end{subfigure}
    \caption{Topologies considered for evaluation}
    \label{fig:initial_topologies}
\end{figure}

We investigate two types of abstract topologies as these were the ones considered by the designers of the network. One is a tree-like topology (figure~\ref{fig:tree_topo}) which connects the consumers while the other adds a few more connections to make it a circular or ring-like topology (figure~\ref{fig:circular_topo}). It is important to note that as this methodology evaluates fairness of a network topology given the demands, we could also use the workflow presented here within a topology optimizer which produces the topologies for evaluation and have a more robust final design of a network.

\subsection{Solvers}
Since the optimization problem with equation~\eqref{eq:modified_cost_function} is a QUBO problem, we can attempt to solve this in a quantum annealer which has been described in more detail in section~\ref{subsec:QC}. We implemented the QUBO matrix formed by equation~\eqref{eq:modified_cost_function}. This matrix is then provided to the qbsolv~\cite{Booth:qbsolv:2017} solver from D-Wave Systems to connect to their QPU machine with the automatic embedding to the system through the EmbeddingComposite~\cite{DWave:git:2017} function.

We also used the METIS~\cite{Karypis:1998:FHQ:305219.305248} graph partitioning solver to compare a classical solution with the one obtained from qbsolv.

\subsection{Workflow}
The workflow to evaluate fairness is shown in algorithm~\ref{algo:workflow}. The results of the workflow output can help inform the user on the fairness of the evaluated topology, $G(V,E)$.

\begin{algorithm}[H]
    \label{algo:workflow}
    \SetAlgoLined
    \SetKwInOut{Input}{input}\SetKwInOut{Output}{output}
    \SetKwData{KPIarray}{kpiArray}
    \SetKwFunction{QUBOSolver}{QUBOSolver}
    \Input{The topology, $G(V,E)$, to be evaluated}
    \Input{ Weighted demands from consumers, $w$}
    \Input{ Maximum number of producers to be considered, $N_k$}
    \BlankLine
    \For{$k \leftarrow 1$ \KwTo $N_k$}{
        $\mathbf{X}_k \leftarrow$ \QUBOSolver($k$, $w$, $G(V,E)$) \tcc*[f]{QUBOSolver solves eq~\eqref{eq:modified_cost_function} using qbsolv or METIS}
        
        \KPIarray[$k$] $\leftarrow$ KPI($\mathbf{X}_k$) \tcc*[f]{KPI evaluations using eq~\eqref{eq:KPI_equation}}
    }
    \Output{\KPIarray}
    \caption{Workflow used to evaluate the topology KPIs}
\end{algorithm}

\section{Results}
\label{Chap:Results}
In this section we provide the results of our proposed topologies on the fairness and overall KPI. We also discuss on the quality of the solutions produced by both solvers.

\begin{figure}
    \centering
    \begin{subfigure}[b]{0.45\textwidth}
        \centering
        \includegraphics[width=0.95\linewidth]{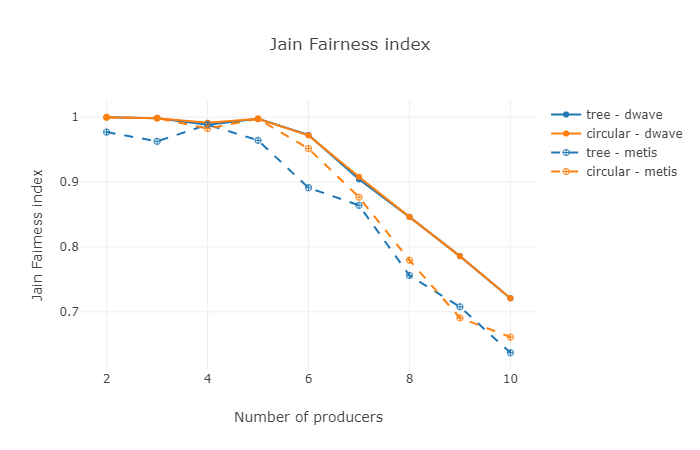}
        \caption{Jain fairness index}
        \label{fig:jain_index}
    \end{subfigure}
    ~ %add desired spacing between images, e. g. ~, \quad, \qquad, \hfill etc. 
    \begin{subfigure}[b]{0.45\textwidth}
        \centering
        \includegraphics[width=0.95\linewidth]{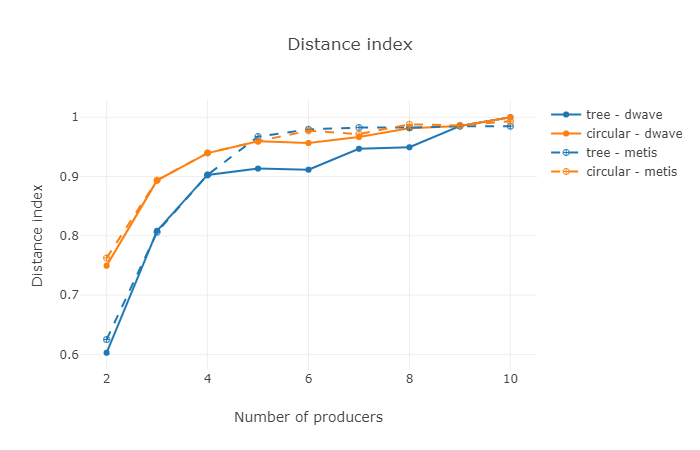}
        \caption{Distance index}
        \label{fig:dist_index}
    \end{subfigure}
    \begin{subfigure}[b]{0.6\textwidth}
        \centering
        \includegraphics[width=0.95\linewidth]{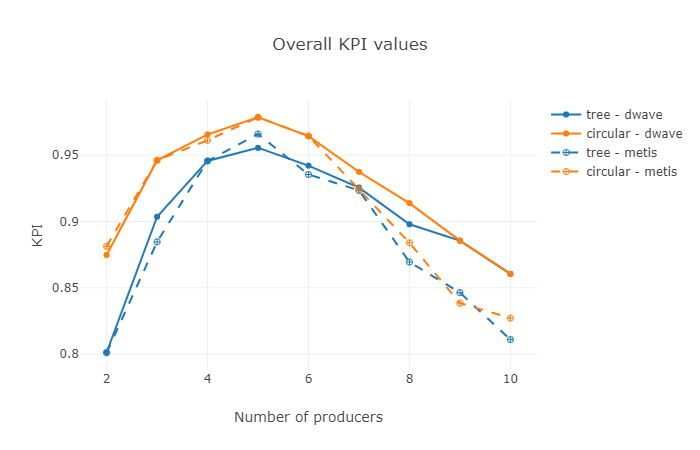}
        \caption{Overall fairness KPI}
        \label{fig:KPI_index}
    \end{subfigure}
    \caption{Fairness indices between multiple topologies, number of producers, and solvers}
    \label{fig:KPI_indices}
\end{figure}

Figure~\ref{fig:KPI_indices} show the different fairness indices when we increase the number of producers as computed by the solvers for different topologies. We notice from the Jain fairness index (fig~\ref{fig:jain_index}) that for all the topologies, the index reduces as we increase the number of producers. As for the distance index (fig~\ref{fig:dist_index}), it indicates that as the number of producers increase in the network, the energy transportation costs reduce, thus increasing the index value. 

The overall KPI indicates the quality of the solution provided by the solvers and we can see from figure~\ref{fig:KPI_index} that the optimum quality of the solution occurs when the number of producers in the network (for both topologies) is approximately 5. As we change the weighting factor $\alpha_j$ of the overall KPI (equation~\eqref{eq:KPI_equation}), we would switch the overall KPI between that of fully fair (fig~\ref{fig:jain_index}) to minimization of cluster distance (fig~\ref{fig:dist_index}).

\begin{figure}
    \centering
    \begin{subfigure}[t]{0.45\textwidth}
        \centering
        \includegraphics[width=0.65\linewidth]{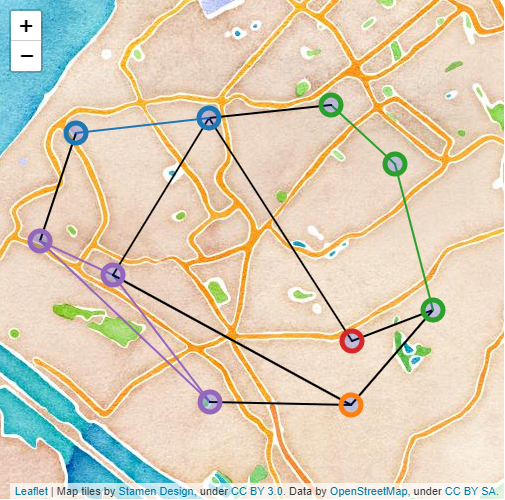}
        \caption{Node assignments for each producer (Map)}
        \label{fig:sample_solution}
    \end{subfigure}
    ~ %add desired spacing between images, e. g. ~, \quad, \qquad, \hfill etc. 
    \begin{subfigure}[t]{0.45\textwidth}
        \includegraphics[width=0.95\linewidth]{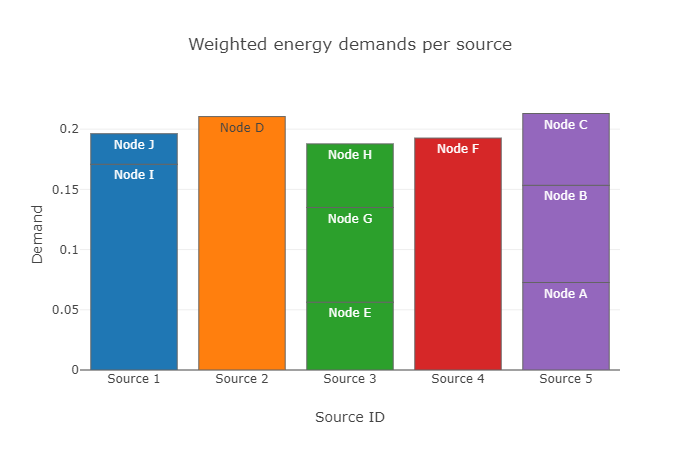}
        \caption{Node assignments for each producer}
        \label{fig:sample_assignment}
    \end{subfigure}
    \caption{qbsolv solution for $k=5$ and a circular type topology}
    \label{fig:sample_dwave_solution}
\end{figure}

\begin{figure}
    \centering
    \begin{subfigure}[t]{0.45\textwidth}
        \centering
        \includegraphics[width=0.65\linewidth]{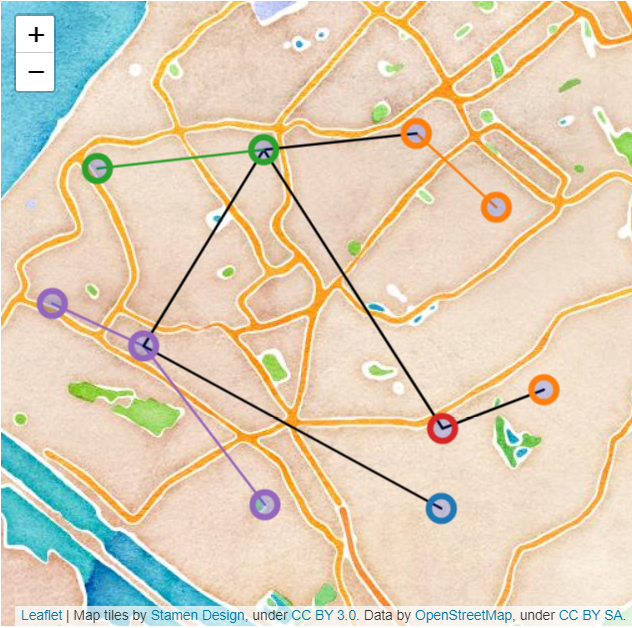}
        \caption{Node assignments for each producer (Map)}
        \label{fig:sample_solution_tree}
    \end{subfigure}
    ~ %add desired spacing between images, e. g. ~, \quad, \qquad, \hfill etc. 
    \begin{subfigure}[t]{0.45\textwidth}
        \includegraphics[width=0.95\linewidth]{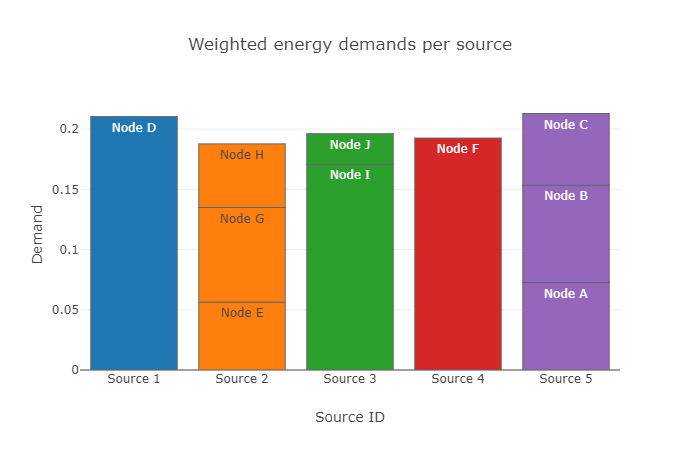}
        \caption{Node assignments for each producer}
        \label{fig:sample_assignment_tree}
    \end{subfigure}
    \caption{qbsolv solution for $k=5$ and a tree type topology}
    \label{fig:sample_dwave_solution_tree}
\end{figure}

Another observation from figure~\ref{fig:KPI_index} is that the circular-like topology produces better quality assignments (figure~\ref{fig:sample_dwave_solution}) compared to the tree-like topology (figure~\ref{fig:sample_dwave_solution_tree}) from both solvers. As for the comparison between the different solvers, we do not notice any significant difference between the solutions produced by both types of solvers, and this is expected since the number of consumer nodes in the network is relatively small, and from \cite{2017arXiv170503082U}, we would not notice much added advantage until we get to networks of size, $n \gtrapprox 200$. However, the same implementation and workflow can be used for more topologies and network sizes in the future.

\newpage
\section{Conclusions}
\label{Chap:Conclusions}
From the previous section, we have seen that a heat grid network can have an optimum number of producers. Adding more producers would reduce the fairness or the overall quality of the assignments. Thus, another conclusion which we can draw from the results is that during the conceptual design phase, the fairness concept should be introduced to help provide a robust network for the future. Trivially, adding more connections between nodes should provide an increased fairness for all the producers within the network. However, this does not take into account the original costs of laying the connections in the first place. We also note that it is up to the user to interpret the KPI and how it can be used to alleviate the monopolistic problem of DHNs. The KPI just serves as a quantifiable measure of fairness which can be tuned.

The workflow can also be adapted to be more time dynamic by performing the fairness optimization step in a quasi-static manner (i.e. by evaluating the solution at each timestep for the nodes instead of currently taking the maximum of the demand profile).

On a technological note, the implementation is suitable to be solved in a quantum annealer through qbsolv, and the advantages could be promising if the network to be evaluated consist of a large number of edges and vertices. We could also implement the same QUBO problem in gated quantum computers and use the QAOA hybrid algorithm to solve the optimization problem.

\section{Acknowledgements}
The author would like to acknowledge the useful discussions he had with his fellow colleagues, Paul Egberts, and Can Tumer at the Heat Transfer and Fluid Dynamics (HTFD) department in TNO, regarding topology optimization and district heating network fairness. The author would also like to acknowledge the support from the department manager, Jeremy Veltin, in allowing for this project to be completed.

\newpage
\bibliographystyle{ACM-Reference-Format}
\bibliography{refs}

\end{document}